\documentclass[12pt]{article}
\usepackage{graphicx}

\input epsf
\def\beq{\begin{eqnarray}}
\def\eeq{\end{eqnarray}}

\def\k{{\bf k}}

\def\lsim{\mathrel{\rlap{\lower3pt\hbox{\hskip0pt$\sim$}}
     \raise1pt\hbox{$<$}}}         
\def\gsim{\mathrel{\rlap{\lower4pt\hbox{\hskip1pt$\sim$}}
     \raise1pt\hbox{$>$}}}         

\usepackage{amsmath}
\usepackage{amsfonts}
\usepackage{verbatim}

\begin{document}

\begin{titlepage}

\thispagestyle{empty}

\begin{flushright}
{NYU-TH-09/04/15}
\end{flushright}
\vskip 0.9cm

\centerline{\Large \bf Quantum Liquid Signatures in  Dwarf Stars}
\vskip 0.2cm
\centerline{\Large \bf }                    

\vskip 0.7cm
\centerline{\large Gregory Gabadadze and David Pirtskhalava}
\vskip 0.3cm
\centerline{\em Center for Cosmology and Particle Physics,  
Department of Physics,}
\centerline{\em New York University, New York, 
NY  10003, USA}

\vskip 1.9cm

\begin{abstract}

We develop further the proposal of arXiv:0806.3692 that 
a new state of matter -- charged condensate of spin-0 nuclei -- 
may exist in helium-core dwarf stars.  The charged condensate  
and its fluctuations are described by an effective field 
theory Lagrangian. The spectrum of bosonic fluctuations  
is gapped, while electrons,  at temperatures of interest,  
give rise to gapless  excitations  near the Fermi surface. 
These properties determine the evolution of the 
dwarfs with  condensed cores. 
In particular, we show that such  dwarf stars  would cool 
significantly faster than their crystallized counterparts. 
As a result, the luminosity  function for the helium-core 
dwarfs will have a sharp drop-off after the condensation. 
It is tempting to interpret the  recently discovered 
abrupt termination of a sequence of 24 helium-core
dwarf candidates in NGC 6397  as a signature of the charged 
condensation.

\end{abstract}

\vspace{3cm}

\end{titlepage}

\newpage

\begin{center}
{\large \bf 1. Introduction and Summary}
\end{center}

\vskip 0.3cm

There is roughly  five orders of magnitude between what may be called the 
atomic scale,  {\it Angstr\"om} $ \sim 10^{-8}~cm$,   and the nuclear scale, 
{\it Fermi} $ \sim 10^{-13}~cm$. If in a neutral system of a large number 
of electrons and  nuclei  average inter-particle  separations 
are between these two scales, then neither atomic  nor nuclear effects 
will play  any significant role. Moreover, the nuclei can be treated as 
point-like particles. 

Such a system of electrons and nuclei  constitutes cores of 
white dwarf stars. Up to a factor of a few, these are 
roughly Earth's size  solar-mass objects; their  
central mass  density  may range over  $\sim (10^6-10^{10})~g/cm^3$, 
most of them being near the lower edge of this interval.
Since  the dwarf stars exhausted  thermonuclear fuel in their cores 
already, they evolve by cooling \cite {Mestel};   
the ones that we consider in this work  cool from $\sim 10^8~K$ 
down to lower temperatures.

At temperatures of interest  the electron de Broglie wavelengths
overlap with each other, and they obey  the 
laws of quantum  statistics. Hence,  the electron properties  
are universal throughout all  the  dwarf stars; their number density 
may range in the interval  $\sim (0.1-5~MeV)^3$, and at temperatures 
of interest they form a degenerate quantum gas 
(their Fermi  energy is greater that their 
interparticle Coulomb interaction energy).

As to the nuclei -- since they're heavier  and  
their de Broglie wavelengths are shorter -- 
they may exhibit different  properties depending on the 
values of density and temperature at hand: they may 
crystallize \cite {MestelRuderman}
as it should be the case for a majority of  dwarf stars, 
or alternatively, may form a quantum liquid \cite {GGRRwd},
when densities are high enough. Identification and qualitative description 
of  quantum liquid signatures  in the cores of high-density 
cool white dwarf stars is a main subject of the present work.

\vspace{0.1cm}

As a typical dwarf star cools down,
the Coulomb interaction  energy in a classical plasma of  
charged nuclei  will significantly  exceed their classical thermal energy, 
and the nuclei, in order to minimize energy,  
would organize themselves into a crystal lattice 
\cite {MestelRuderman}. In most of these  cases  
quantum effects of the  nuclei should be negligible;  
for instance, the Debye temperature should  be less than the 
temperature at which crystallization takes place, and  
the de Broglie wavelengths of the nuclei should be much 
smaller than  the average internuclear separations. 
This indeed is the case in majority of white dwarf stars,  
the cores of which  are  composed of carbon and/or oxygen nuclei 
and span the interval of mass densities around $\sim ( 10^6-10^8) ~g/cm^3$.

However, there exists a class of dwarf stars in which the nuclei
enter the quantum regime before the above-described classical 
crystallization process sets in \cite {Ashcroft,Chabrier}.
Among these, furthermore, there is a relatively small 
subclass of the dwarf stars for which the temperature $T_c$,   
at which the de Broglie wavelengths of the nuclei 
begin to overlap, is  higher than the would-be 
crystallization temperature $T_{cryst}$ (see \cite {GGRRwd} and 
discussions in Section 2 below). 
Then, right below $T_c$,  the quantum-mechanical uncertainty in the 
position of the charged nuclei is greater that the average inter-nuclear 
separation. This is diametrically opposite to the crystallized state  
where the nuclei would have well-localized positions with slight 
quantum-mechanical fuzziness due to their zero-point oscillations.

What is then an adequate description of such a  state? 
It was argued in \cite {GGRRwd,GGRR1}  that such a system, instead 
of forming a crystalline lattice at  $T_{cryst}$,  could condense 
at $T_c>T_{cryst}$, owing to the quantum-mechanical probabilistic 
``attraction'' of Bose particles to occupy one and the same zero-momentum 
state, and  leading to  a quantum liquid in which the 
charged spin-0 nuclei  would form a macroscopic quantum  state 
with a large occupation number -- the charged condensate. 

It was proposed in \cite {GGRRwd} that the conditions in 
certain high-density helium-core  white dwarfs  (He WDs) 
are appropriate to form such  
charged condensate.  Here  we will argue that similar 
effects could  take place in lower-density He WDs, which are 
more relevant for observations \cite {Shapiro,Eis,24}, 
as well as in superdense ($\sim 10^{10}~g/cm^3$)  
carbon-core white dwarfs (C WDs).

The observational consequences of the charged condensate in astrophysical 
objects can be significant.  The bosonic part of the liquid 
is superconducting  as its  spectrum of  long-wavelength  
fluctuations exhibits a mas gap; hence  
these quasi-particles contribute to  the specific heat 
of the substance only in an exponentially suppressed  
way \cite {GGRRwd}. This affects the cooling rate for dwarf stars.
In particular, we will show that the He WDs  with the condensed cores  
would cool much faster than the ones in which 
condensation does not take place.

We should emphasize though that a careful consideration of 
white dwarf cooling rates requires detailed account of 
composition and dynamics of their envelopes, which is 
a complex and less certainly known subject (for detailed 
discussions see, e.g., \cite {AntonaRev,LibertRev}). We will not 
attempt here to enter these studies.  Instead, we consider an 
over-simplified model just  to emphasize our main point that 
if the charged condensation takes place in the cores of white dwarfs, 
they'll necessarily cool faster, and there will be a drop-off in 
their luminosity function. 

One may wonder whether the effects that  we're discussing can be 
entirely obfuscated by the  uncertainties in the envelope composition. 
Although this may well  be 
the case for cooling times, there is nevertheless a model independent 
prediction of the charged condensation: the luminosity function
will have a sharp drop-off after the condensation, with subsequent 
growth governed by a  shallower slope. 

The recently  discovered and studied  in Ref. \cite {24} 
24 helium-core dwarf candidates in NGC 6397  
exhibit a termination of the sequence at low luminosities. 
It is tempting to attribute this to the drop-off due to the 
charged condensation. Whether this proposal can withstand 
more detailed scrutiny remains to be seen.

The organization of the paper is as follows: In Section 2
we give qualitative arguments why condensation may be preferred over
crystallization in certain WDs. In Section 3 we summarize and further develop
the effective field theory approach of \cite {GGRRwd,GGRReff} to charged 
condensation, and describe the spectrum  that is relevant 
for cooling of WDs. In Section 4 we calculate cooling times 
for He WDs and compare them with those of core-crystallized WDs. 
Last but not least,  we discuss the drop-off of the luminosity 
function after charged condensation.

\vspace{0.3cm} 

\begin{center}
{\large \bf 2. Condensation versus Crystallization}
\end{center}

\vskip 0.3cm

We concentrate on WDs that below certain temperature  
cool by releasing stored in them heat from their surface. 
At the beginning of that stage  the ions (nuclei) 
form a classical Bose gas that cools with  constant 
specific heat  (the so-called Mestel cooling 
\cite {Mestel}) down to some  temperature 
$T_{\text{cryst}}$, at which Coulomb repulsion starts to dominate 
by about two orders of magnitude over the average thermal energy, 
and the core undergoes crystallization transition \cite {MestelRuderman}.
The crystallization temperature in the classical regime is quantified 
by the ratio
\beq
\Gamma\equiv\frac{E_{\text{Coulomb}}}
{2E_{\text{Thermal}}/3}=\frac{(Ze)^2/4\pi d}{k_BT}\,,
\label{Gamma}
\eeq
where $e$ is the electric charge, 
$Ze$ is the charge of a nucleus, $k_B$ denotes the Boltzmann constant, 
$J_0$ is the electron number-density, $d\equiv (3Z/4\pi J_0)^{1/3}$ is 
the average inter-ion separation, while $T$ denotes the core temperature 
of the star. The ion plasma is expected to undergo  crystallization
once the temperature  drops to a value for which $\Gamma\simeq 180$
\cite {vanHorn,Ichi,DeWitt}.

The above arguments are entirely {\it classical}. 
The temperature scale that  determines the 
classical versus quantum nature of the 
crystallization transition is the Debye temperature
\begin{equation}
\theta_D\equiv {\hbar\Omega_p \over k_B },\qquad \Omega_p=
\left (\frac{J_H (Ze)^2}{m_H} \right )^{\frac{1}{2}},
\label{Debye}
\end{equation}
where $\Omega_p$ is the plasma frequency of the ion gas, 
$J_H=J_0/Z$ denotes the ion number density and $m_H$ denotes 
the mass of a single ion (the subscript $H$ 
stands for ``heavy''). Up to a factor of $\sqrt{3}$, $\Omega_p$ 
coincides with the frequency of zero-point oscillations
of the ions in crystal sites,  $w_0=\Omega_p/\sqrt{3}$.

The white dwarfs with  $T_{\text{cryst}}>\theta_D$  cool according to 
the above-described classical scenario.  Often however, $\theta_D$ may 
significantly exceed  $T_{\text{cryst}}$ \cite{Ashcroft}. 
In such a case, quantum zero-point oscillations 
should be taken into account in order to derive the crystallization 
temperature.  This seems to delay the formation of quantum crystal, lowering 
$T_{\text{cryst}}$ from its classical value at most by about 
$\sim 10 \% $ \cite{Chabrier}.   Since this is a small change for  
the estimates that we're after here,  
we will consider the classical  value of $T_{\text{cryst}}$  
to be a good approximation, even in the quantum case, keeping in mind 
that $T_{\text{cryst}}$ may overestimate  somewhat the crystallization 
temperature of the substance.

\vspace{0.1in}

However, there exists a third and very important temperature scale, 
relevant for studying the cooling of white dwarf interiors. 
It is the ``critical''  temperature  $T_c$, at which the de 
Broglie wavelengths of the ions start to overlap
\begin{equation}
T_c\simeq\frac{4\pi^2}{3m_Hd^2}\,.
\label{Tc}
\end{equation}
Below $T_c$ quantum-mechanical uncertainties in the ion positions become 
greater than  an average inter-ion separation. Hence the latter concept 
looses its meaning as a microscopic characteristic  of the system, and  
the ions enter a quantum-mechanical regime of indistinguishability\footnote
{The de Broglie wavelength above is defined as $\lambda_{dB}=2\pi/|\k|$,
where $\k^2/2m_H = 3k_B T/2$. We define $T_c$ as the temperature at which 
 $\lambda_{dB}\simeq d$. Note that this differs by a numerical 
factor from the standard definition of the {\it thermal} de Broglie 
wavelength,  $\Lambda \equiv \sqrt{ 2\pi /mk_BT}$,  that appears  
as a natural scale in the partition function. See comments on the 
rationale for our choice of $T_c$ below.}. Below $T_c$ the wavefunction of 
the many-body system of spin-0 ions should be symmetrized, and this would 
unavoidably lead to probabilistic ``attraction''  of the bosons to 
occupy the same quantum state.

Therefore, when  crystallization temperature $T_{\text{cryst}}$
is lower than  $T_c$, the system  may  instead  
undergo condensation  into a macroscopic zero-momentum 
quantum state  with  large occupation number -- 
the charged condensate.   

Once in the condensate, the boson positions are entirely uncertain  
while their momenta equal to zero. In order for such a system  
to crystallize later on, each of the bosons should acquire the momentum 
determined by  the zero-point energy  of the 
crystal ions, $\k_0^2= 2m_H \Omega_p/\sqrt{3}$. The latter  
is greater than the Fermi momentum, as well as typical momenta of
fermionic quasiparticles. Hence, the fermions will not be able to 
transmit to the condensed bosons momenta comparable with  $|\k_0|$.
Therefore, the transition to the crystallized state can only happen 
spontaneously.  Such a transition could 
take place  as it would lower the energy of the entire system  
due to the favorable electrostatic screening. However,  given that the 
spectrum of bosonic quasiparticles is gapped, the process 
will be exponentially suppressed at temperatures  
below the gap scale. Hence, the condensate should not be 
expected to  undergo subsequent crystallization, at least for 
a long period of time.

Another crucial question is whether the expression for the critical 
temperature  (\ref {Tc}) gives an 
accurate estimate for the actual condensation temperature 
$T_{\rm cond}$ at which the phase transition 
would take  place. If we were to deal with a {\it non-interacting} 
system of  Bosons then the known BE condensation temperature, 
$T^{BE}_{\rm cond} \simeq (1.27/m_Hd^2)$,  
would have been an order of magnitude smaller  than  what the estimate  
(\ref {Tc})  suggests.   However, it has been known that already 
{\it weak repulsive} interactions between bosons increase  
the condensation temperature; this is  consistent with ones expectation 
that the repulsion makes easier for the condensation in the momentum space 
to take place (indeed, the BE condensation is a condensation in the 
momentum space, while the coordinate space  wave-functions are entirely 
delocalized) see, e.g., \cite {Huang} and references therein.

In the case of weakly interacting bosons the increase of the 
condensation temperature is small since the 
interactions are weak.  In our case, however,  interactions 
between spin-0 nuclei are strong, in a sense of a  
many-body system, as we are about to argue below. In this case  
we  would expect $T_{\rm cond}\gg T^{BE}_{\rm cond}$. 
Since we have no means to evaluate $T_{\rm cond}$
accurately, we use the  expression (\ref {Tc}) as a reasonable and 
physically motivated approximation  for the interacting system 
$T_c\sim T_{\rm cond}$.  To this end, 
the criterium that we adopt for the condensation to take place is 
\beq
T_c \gsim (a~ few)~ T_{cryst}.
\label{crit} 
\eeq

\vspace{0.1in}

The charged condensate is somewhat similar, 
but also  differs by its strong coupling,  from the 
Bose-Einstein  condensate of charged spin-0 particles. 
Analytic studies of a BE  condensate of charged scalars 
in the past (see, e.g., Ref.  \cite{Foldy} and references citing it) 
relied on a small departure from the condensation  of free particles,    
and used an expansion in a parameter $r_s$ 
that is defined as the ratio of the average interparticle separation to 
the would-be Bohr radius for the  boson.  For a weakly non-ideal 
system of bosons the $r_s$-ratio 
is small $r_s\ll 1$, and the ground state is the  
weakly-non-ideal BE condensate. Furthermore, as zero-temperature numerical 
simulations show (see, e.g., \cite {Alder,It}, and 
references therein), for $1\lsim r_s\lsim 160$ the condensate and 
crystal state begin  to coexist. Furthermore, for  $r_s\gsim 160$ 
the entirely crystallized state is  a ground state.  Presumably, the 
finite temperature effects would increase  the  value of  $r_s$ at which  
the crystallization takes place  only by a factor of a few, but not by 
an order of magnitude (see, e.g., \cite {ChabrierT}). 

In the system of nuclei and electrons the $r_s$-ratio 
cannot be made  small without entering the regime where 
nuclear interactions become dominant. Instead, in the case considered 
in \cite {GGRRwd}, $r_s\sim(10-100)$.  In the present work we will consider
even larger values, $r_s \sim 10^3$, which are relevant for 
observed He WDs. Yet  we argue that because the 
temperature $T_c$ at which the de Broglie wavelengths of bosons  
start to overlap is greater than the crystallization temperature, the 
system, upon gradual cooling,  should  settle in  the 
charged condensate state due to the quantum-mechanical 
probabilistic properties of the indistinguishable bosons. 
In fact, our criterium $T_c>T_{cryst}$  implies that the charged 
condensation could take place at fine temperature $T_c$ 
for $r_s < 2400$. It is also worth pointing out  in this regard that 
the expansion parameter in our case ends up being 
$1/r_s$,  as long as the system is described in an effective 
Lagrangian approach.

The seeming contradiction with the numerical results may  be reconciled by  
the fact that the charged condensate can  only be a metastable state 
\cite {GGRRMeta}, while the crystal should arguably  be the  lowest 
energy  state for  $r_s \gg 160$. Most of the numerical 
simulations use the test wavefunction approach that minimizes the energy; 
finding a long-lived state which represents only a local minimum may not 
be easy in this  approach\footnote{It is interesting to note that in Ref. 
\cite {Alder} a metastable quantum liquid branch was observed for  
$r_s \gsim 160$.  At this stage it is hard to speculate whether 
this is the branch that we're  discussing  here.}. 

One  check  of this proposal is that  the small 
fluctuations of the charged condensate  have no unstable modes, 
suggesting that it represents at least a local minimum. Also, these 
fluctuations are rather different from those of a would-be crystal, 
ordinary clod plasma, or weakly-coupled BE condensate. 
For discussions of non-perturbative  (meta)stability  of the 
charged condensate in a different context, see \cite {GGRRMeta}.

\vspace{0.3cm} 

\begin{center}
{\large \bf 3.  Charged Condensate and its Fluctuations}
\end{center}

\vskip 0.3cm

We use an effective  Lagrangian description 
to study  the charged condensate\footnote{An effective field 
theory that describes  the charged condensate  was discussed  
in Refs. \cite {GGRRwd,GGRR1}. In this section we briefly describe 
and expand some of the  results of these works.}. 
We focus on the  zero-temperature limit, even though 
realistic temperatures in He WDs are well above zero
(for calculations of the finite temperature effects, 
see \cite {Dolgov}).  The validity of the zero-temperature approximation is 
justified  {\it a posteriori} and goes as follows:  the spin-0 
nuclei undergo the condensation to the zero-momentum state;  
while they do so  they cannot excite their own phonons since the latter 
are gapped with the magnitude  of the gap being greater  than 
the condensation temperature. On the other hand, 
the condensing charged bosons can  and will  excite 
thermal fluctuations in the  fermionic sector that is gap-less. 
Therefore, all the thermal fluctuations 
will end up being stored in the  fermionic quasiparticles near the Fermi 
surface. For the later, however, the finite temperature effects aren't 
significant since their Fermi energy is so much higher,  
$T/J_0^{1/3}\ll 10^{-2}$.

After these comments we turn to the effective Lagrangian. 
By $m_H$ we denote the mass of a (heavy) nucleus of charge $Ze$ and 
atomic number $A$ (helium-4, carbon, oxygen), by $\mu_f$ the electron 
chemical  potential, and  by $m_e$ the electron mass. The   
following  hierarchy of scales, $m_H \gg {\rm max}[\mu_f,  m_e]$, 
is a starting point for the effective Lagrangian construction. 

We begin  at scales that are well below the heavy mass scale
$m_H$, but  somewhat above the scale set by 
${\rm max}[\mu_f,  m_e]$. Hence the electrons are described  
by their Dirac Lagrangian, while for the  description of 
the nuclei we will use a charged scalar {\it order parameter} $\Phi(x)$.
As it was shown in \cite {GGRReff}, in a non-relativistic  
approximation for the nuclei,  an effective  Lagrangian proposed by  
Greiter, Wilczek and Witten  (GWW) \cite {GWW} in 
a context of superconductivity, is also applicable for  
the description of  the charged condensation, given that  an appropriate 
reinterpretation of its variables and  
parameters is made.

The construction of the  GWW effective Lagrangian  is based on the following  
fundamental principles: it is  consistent  with the translational, 
rotational, Galilean and the global $U(1)$ symmetries, 
preserves the algebraic relation  between the charged current density 
and momentum density, gives the Schr\"odinger equation for the 
order parameter in  the lowest order, and is gauge invariant \cite {GWW}.  
Combined  with  the electron dynamics the GWW effective Lagrangian reads 
(we omit for simplicity the standard Maxwell term):
\beq
{\cal L}_{eff} = {\cal P} \left (  
 {i\over 2} ( \Phi^*  D_0 \Phi -  (D_0 \Phi)^* \Phi)-
{| D_j  \Phi|^2  \over 2m_H} \right )\,+
{\bar \psi}(i\gamma^\mu D^f_\mu -m_f)\psi,
\label{Leff}
\eeq
where we use the standard notations for covariant derivatives with the 
appropriate charge assignments: $D_0 \equiv (\partial_0  - iZe A_0)$, 
$ D_j \equiv ( \partial_j - i Ze A_j) $,
$D^f_\mu = \partial_\mu +ie A_\mu $, 
while  ${\cal P}(x)$ stands for a general polynomial function of its argument
\footnote{In a more complete treatment one should also 
add to the Lagrangian terms $\mu_{NR} \Phi^*\Phi$,  
$\lambda (\Phi^*\Phi)^2/m_H^2$, and the higher dimensional operators
that are consistent 
with all the symmetries and conditions that lead to (\ref {Leff}).
Here $\mu_{NR}$ denotes a non-relativistic chemical potential for 
the scalars.  These terms will not be important for 
the low-temperature spectrum of small perturbations  we're interested in, 
as long as  $\lambda\lsim 1$ and  $J_0\ll m^3_H$.  
However, near  the phase transition point it is the sign of $\mu_{NR}$ that 
would  distinguish between the broken and symmetric phases,  so 
these terms should be included for the discussion of the symmetry 
restoration.  We also note that the scalar part of  (\ref {Leff}) is somewhat 
similar to the Ginzburg-Landau (GL) Lagrangian for superconductivity. However, 
there is a significant difference between the two. The coherence length 
in the GL theory is many orders of magnitude greater than the  
average interelectron separation, while in the present case, the ``size of the 
scalar''  $\Phi$ is smaller that the average interparticle 
distance.}. 

The coefficients of this polynomial, 
${\cal P}(x) = \sum^{\infty}_{n=0} c_n (x^n / \Lambda^{4n})$,
are dimensionful numbers that are inversely proportional to powers of 
a short-distance  cutoff of the effective  field theory.

Once the basic Lagrangian is fixed,  we  introduce  the 
electron chemical potential term $\mu_{f} \psi^+ \psi\,$, 
where  $\mu_{f}=\epsilon_F=[(3 \pi^2 J_0)^{2/3}+m_f^2]^{1/2}$. This gives 
a nonzero electron number density $J_0$ which is related to the 
Fermi momentum $k_F= (3 \pi^2 J_0)^{1/3}$.  This is also the only term that  
at the tree level sets a frame in which the electron total 
momentum is zero. The quantum loop corrections due to this term 
will generate additional Lorentz-violating terms in the bosonic sector 
of the theory \cite {GGRRwd,Dolgov,GGRReff}.
 
There exists  a solution of  the equations of motion 
that follow from the effective Lagrangian (\ref {Leff}). 
This solution takes the form  \cite {GGRRwd}:
\beq
Z|\Phi|^2 = J_0\,,~~~ A_\mu=0,~~~~{\cal P}^{\prime}(0)=1\,.
\label{sol}
\eeq 
(We use  the unitary gauge $\Phi = |\Phi|$).
The condition ${\cal P}^{\prime}(0)=1$ is satisfied by any 
polynomial functions  ${\cal P} (x)$ for which the first coefficient is 
normalized to unity
\beq
{\cal P} (x) = x + C_2 x^2+...\,.
\label{px}
\eeq
The above solution describes a neutral system of 
negatively  charged electrons  of charge density 
$- eJ_0$,  and positively  charged scalar (helium-4 nuclei) 
condensate  of charge density $Ze\Phi^+\Phi=eJ_0$.
This describes the  
condensate and not  a standard crystal in a  long 
wavelength  approximation where lattice inhomogeneities 
can be neglected, or ordinary 
cold plasma. This becomes more clear after one calculates the spectrum 
of small perturbations about the  homogeneous solution and finds that it is 
rather different from the  spectra  of nearly free BE condensate, 
crystal lattice vibrations, or from plasma fluctuations.

Calculation of the spectrum of small perturbations 
is straightforward (here we follow conventions of  
\cite {GGRReff}).  There are two transverse polarizations 
of a massive photon which propagate with the conventional massive 
dispersion relation 
\beq
\omega^2 = \k^2 +m_\gamma^2, ~~~~m_\gamma^2\equiv {Z e^2 J_0 \over m_H}\,. 
\label{magnetic}
\eeq
Moreover, there is one longitudinal mode (phonon) with the following 
unconventional dispersion relation 
\beq
\omega^2 \simeq m_\gamma^2\left [1 + {\k^4 \over \k^2 m_0^2 + 4M^4 }\right ]
\simeq  m_\gamma^2 +  {\k^4 \over 4m_H^2 }\,.
\label{electric}
\eeq
Here $M\equiv \sqrt{m_Hm_\gamma}$,  
$m_0^2 \equiv m_\gamma^2+ \delta m^2 $, where $\delta m^2\propto 
e^2J_0/E_F$ is the Debye mass squared. As we see, 
$ \omega^2 (|\k|=0)=m_\gamma^2$, and  the bosonic collective 
excitations are gapped.  Hence, the bosonic part of the whole 
system represents a superconducting component, while the fermions, 
at temperatures of interest, can be regarded as the  ``normal'' (dissipative) 
component of the quantum liquid (see more below).

We point out that in the  approximation used above there 
does not appear to be a ${\cal O} (\k^2)$ term on the r.h.s. of the dispersion 
relation (\ref {electric}).  Such a  term emerges (see \cite {GGRReff}) in 
the $1/m_H^3$ order of the heavy mass expansion,  and is proportional 
to $- m_\gamma^2 \k^2/m_H^2$.
This term is responsible for the ``roton-like'' behavior of 
small fluctuations discussed in \cite  {GGRReff}. 
The dispersion relation (\ref {electric}) is 
applicable for $k^2\gsim m_\gamma^2$, which corresponds to distances 
shorter than $1/m_\gamma$ -- the scale that encompasses a huge 
number of particles.

Note also that the phonon dispersion 
relation (\ref {electric}) is different from that of a conventional 
phonon in  a crystal for which $\omega \propto  |\k|$,  near the origin.
Hence,  the substance that we're discussing is different from 
a crystal, and it is also different from classical plasma.  
The following line of arguments also suggests that 
the charged condensate differs from the ordinary 
BE condensate of nearly-free bosons too:  In the limit of switched off 
interactions,  $e\to 0$,  the dispersion relation (\ref {electric})
reduces to $\omega = \k^2 /2m_H$. This is nothing but a dispersion 
relation for  a lowest excitation in a  BE condensate of  {\it free} 
bosons of mass $m_H$.  However,  for realistic 
values of the parameters the second (momentum dependent) 
term in (\ref {electric})  is sub-dominant, suggesting that 
the charged condensate has a significant departure from the 
BE condensate of nearly-free bosons.

What is important for the present  work is that the bosonic 
collective excitations give rise to exponentially suppressed  
contributions to the value of specific heat of this substance.
Typical suppression scales as 
${\rm exp} (-m_\gamma /T)$, and since  for the He WDs  
$m_\gamma \simeq 3~KeV$, 
at temperatures below $10^6~K$, these contributions are proportional to 
$\sim {\rm exp} (-30)$ and can be neglected.

This is not the case in dwarf starts in which the nuclei formed a
crystalline  structure. There, the dominant contribution
to the specific heat comes from a crystal phonon.
The latter has a linear dispersion relation $\omega \propto  |\k|$,
and its contribution to the specific heat scales  
with temperature as $T^3$. 

Therefore, to a good approximation, there is 
no phonon contribution in the charged condensate case, while 
it is present when  the cores crystallize. 

As to the electrons, their behavior is similar in  both crystal 
and condensate  cases. At temperatures of interest  they form a 
degenerate Fermi gas  with gap-less excitations near the Fermi surface. 
Their contribution to the specific heat scales
linearly with temperature. In the case of crystallized cores  
this is sub-dominant to the specific heat due to the crystal phonon. 
For the charged condensate, on the other hand, the fermions are the 
dominant contributors to  the specific heat. These properties 
make a difference in cooling of dwarf starts, as we'll discuss 
in detail in the next section\footnote{In the charged condensate  
Cooper pairs of electrons can also be formed,  however, the 
corresponding transition temperature, and 
the magnitude of the gap, are  suppressed by a factor 
${\rm exp} (-1/e_{eff}^2)$, where $e_{eff}^2$  is proportional to 
the value of the inter-electron potential that contains both screened 
Coulomb and phonon exchange. The fact that this potential has attractive 
domain, but is  very  small,   is suggested by  
the static potential found in   
\cite {GGRReff} (see eq. (\ref {potential}) below); 
the latter is down by a power of 
a large scale $M$. In other words, the static potential and the 
zero-zero component of the propagator are both 
suppressed as $D_{00}\sim 1/r_s\sim (10^{-3}-10^{-2})$, 
where $r_s$ was discussed in Section 2. 
Moreover, taking into account the frequency dependence  
via the Eliashberg equation does not seem to
change qualitatively the conclusion on a strong  
suppression of the pairing temperature.  

Hence, even though the bosonic  
sector (condensed nuclei) is superconducting at  reasonably high 
temperatures $\lsim 10^{6}~K$, 
interactions with gap-less fermions could dissipate the superconducting 
currents. Only at extremely low temperatures, exponentially close 
to the absolute zero, the electrons could also form a  gap leading 
to superconductivity of the whole system. In the present work we 
consider temperatures at which electrons are not  condensed into  
Cooper pairs, and ignore the finite temperature effects.\label{footnote}}.

\vspace{0.1in}

Finally, let us discuss briefly the  question of impurity
(hydrogen, helium-3, etc.) nuclei that may be present in the 
cores of white dwarfs.  The static potential between two impurity nuclei 
of charge  $Q_1$ and $Q_2$ consists of two parts \cite {GGRReff}: 
\beq
V_{stat}= \alpha_{\rm em}  {Q_1Q_2} \left (  { e^{-Mr}\over \,r}
{\rm cos} (Mr)\, + {4 \alpha_{\rm em} 
\over \pi} { k_F^5{\rm sin}(2k_Fr)\over M^8r^4}\right )\,. 
\label{potential}
\eeq
The first, exponentially suppressed term modulated by a periodic function, 
is due to cancellation between the screened Coulomb potential and 
that of a phonon. The second term, which exhibits a  power-like 
behavior modulated by a periodic function, is due to the 
existence of gap-less excitations near the Fermi surface.  This gives  a 
generalization of the  Friedel potential (see, e.g., \cite {Fetter}) 
to  the case when in addition to 
the fermionic excitations  the  collective modes
associated with  the  charged condensate are also taken 
into account\footnote{Note that for 
spin-dependent interactions the same effects of the charged condensate 
would give a generalization of the  Ruderman-Kittel-Kasuya-Yosida 
(RKKY) potential  \cite {RKKY}.}. As we see, the potential is not 
sign-definite.  It can  give rise to 
attraction between like  charges;  this attraction is due to  
collective excitations of both  fermionic and bosonic degrees of freedom.  
This represents a generalization of the Kohn-Luttinger \cite {Kohn} 
effect to the case where on top of  the fermionic excitations  
the  collective modes of the  charged condensate 
are also contributing (for the discussion of associated 
superconductivity, see  footnote \ref {footnote}). 

For the physical conditions present in  dwarf stars, 
the second term in (\ref {potential}) 
is dominating. This term is  strongly suppressed because of the phonon 
subtraction \cite {GGRReff}.  Thus, attraction between like 
charges due to this potential could give Cooper pairing between 
impurities (even if their  concentration was significant enough) only 
at extremely small temperatures (see footnote \ref {footnote}). 

\vspace{0.3cm} 

\begin{center}
{\large \bf 3. White Dwarf Cooling}
\end{center}

\vskip 0.3cm


Most common dwarf stars have masses approximately equal to 
half of the mass of the Sun, central densities $\sim (10^6- 10^8)~g/cm^3$ 
and  are composed mainly of carbon and/or oxygen. 
Above the largest of the three temperature scales -- 
$T_{\text{cryst}}$, $\theta_D$ and $T_c$ --  the cooling 
process is determined by thermodynamics of the classical 
Bose gas of the ions.  For lower temperatures 
however, the state of the star is determined by the relative 
magnitude of these scales. 

For present purposes, it is more convenient to rewrite the 
expressions for the temperature scales in terms of  the 
\emph{mass} density  $\rho$ measured in $g/cm^3$
\begin{equation}
T_{\text{cryst}}\simeq 10^3\rho^{\frac{1}{3}}Z^{\frac{5}{3}}~K, 
\quad \theta_D\simeq4\cdot10^3\rho^{\frac{1}{2}}~K,~~~~~ T_c =3.5\cdot
10^2\rho^{\frac{2}{3}}/Z^{\frac{5}{3}}~K\,,
\label{Tcryst}
\end{equation}
where the baryon number of an ion was assumed to equal twice 
the number of protons ($A=2Z$) and $\Gamma\simeq180$ was set.

\subsubsection*{Helium White Dwarfs}

White dwarfs composed of helium constitute a smaller sub-class of 
dwarf stars (see, \cite {24,Eis} are references therein). 
They  exhibit best conditions for the charged  condensation. 
Most of helium dwarfs are believed to be formed in binary systems, where 
the removal of the envelope  off  the dwarf 
progenitor red giant by its binary companion happened 
before helium ignition, producing a remnant that evolves to 
a white dwarf with a helium core.
Other astrophysical mechanisms for formation of isolated  
helium white dwarfs  may also be possible \cite {Castellani}. 
In any event, helium dwarf masses range from $\sim 0.5~M_\odot$ 
down to as low as $(0.18-0.19)~M_\odot$, while their envelopes are 
mainly composed of hydrogen. In this work we'll only consider He WDs 
whose hydrogen envelopes are thin, where no thermonuclear reactions 
are taking place. Such WDs cool by radiating 
off the stored in them heat.

Following \cite{Shapiro}, we will consider an over-simplified model of 
white dwarf cooling. Our treatment of what actually is an 
involved process, with significant uncertainties  due to  the envelope
composition and opacity, should not be expected to give quantitatively
precise predictions. Nevertheless, our approach is 
good enough, as we'll see,  to capture the main difference that arises 
in  cooling of dwarfs with the condensed cores.

For definiteness, we consider cooling of a reference helium star 
of mass $M=0.5~M_\odot$ with the 
atmospheric mass fractions of the hydrogen, and heavy elements 
(metallicity) respectively equal to
\beq
 X\simeq 0.99, \quad  \quad Z_m \simeq (0.0002-0.002)~.
\label{eq01}
\eeq
The lower value of the metallicity $ Z_m \simeq 0.0002$ 
is appropriate for the recently discovered 24 He WDs in 
NGC 6397 \cite {24}, but for completeness, we consider a 
wider range for this parameter. Table 1 illustrates different physical 
characteristics of such a star with the mass density $\sim  10^6~g/cm^3$.

The equation of state of a white dwarf interior is  well modeled by 
the polytropes \cite {Shapiro}.  The average density of 
a non-relativistically degenerate star -- which the helium dwarfs 
are an example of -- may be about 5 times  less than the central density. 
For the sake of simplicity, we will 
neglect the nontrivial density profile when dealing with 
thermodynamics of the dwarf stars and consider 
their average and central densities to be roughly equal. 
This won't lead to a significant error in the cooling analysis 
due to the fact  that the variation of the mass density is not dramatic 
in the part of a star which encompasses most of its mass. 

As seen from Table 1, critical temperature for the reference star 
significantly exceeds the crystallization temperature, while Debye 
temperature exceeds both. This star provides appropriate conditions 
for the formation of the charged condensate in its core.

\begin{table}[ht]
\caption{Values of physical quantities for a reference helium white dwarf}
\vspace{0.1cm}
\centering
\begin{tabular}{|c| c|}
\hline\hline
Physical quantity  & Numerical value \\ [0.5ex]
\hline
Electron number density & $(0.13~MeV)^3$ \\
Mass density     & $10^6~g/cm^3$       \\
Separation between atomic nuclei & $  10^{3}~fm$ \\
Crystallization temperature   & $3 \cdot 10^5~K$ \\
Debye temperature  & $4\cdot 10^6~K$ \\
Critical temperature  & $ 10^{6}~K$ \\ [1ex]
\hline
\end{tabular}
\end{table}

In the Appendix we give a  brief summary  of the derivation of the 
luminosity-core  temperature relation for white dwarfs (for details 
see, e.g., \cite {Shapiro}).  The definition of luminosity 
of a cooling star, combined with equation (\ref{eq6}) of the Appendix, 
gives the expression for the cooling rate:   
\beq
-c_v\frac{dT}{dt}=L=CAm_uT^{3.5}.
\label{eq8}
\eeq
Here $c_v$ denotes the specific heat per ion, $m_u$ is the atomic 
mass unit and the constant $C$ is inversely proportional to 
the atmospheric (envelope)  metallicity of a star.  In general, 
variations in  atmospheric  composition could change the cooling 
age significantly.  Although the subject of atmospheric opacities 
is involved, the Kramer's approximation 
endowed with the atmospheric composition given in  (\ref{eq01}) is 
good enough for our purposes of comparing cooling rates of 
crystallized and condensed dwarfs.

The expressions for specific heats of  different components 
of dwarf cores in corresponding  regimes are given in Table 2.

\begin{table}[ht]
\caption{Specific heats of different components of a white dwarf core}
\vspace{0.1cm}
\centering
\begin{tabular}{|c| c|}
\hline\hline
State & Specific heat   \\ [0.5ex]
\hline
Classical Bose gas & $\frac{3}{2}k_B$  \\
Quantum crystal & $\frac{16\pi^4}{5}(\frac{T}{\theta_D})^3 k_B$  \\
Nonrelativistic Fermi Gas & $\frac{\pi^2}{2}\frac{k_BT}{E_F}k_B$  \\ 
Relativistic Fermi Gas & $\frac{(3\pi^2)^{\frac{2}{3}}}{3}\frac{k_B T}
{J_0^{1/3}}k_B$ \\ [1ex]
\hline
\end{tabular}
\end{table}
To quantify the effects of the charged condensation on the cooling rate, 
we consider the ratio of cooling times for two identical helium dwarfs  
with and without the charged condensate in their interiors. 

Integrating (\ref {eq8}),  we find the following expression for the 
cooling time of a star in the Mestel regime 
\beq
t_{He}=\frac{k_B}{CAm_u}\left [ \frac{3}{5}(T_f^{-\frac{5}{2}}-
T_0^{-\frac{5}{2}})+Z\frac{\pi^2}{3}\frac{k_B}{E_F}
(T_f^{-\frac{3}{2}}-T_0^{-\frac{3}{2}})\right ],
\label{eq03}
\eeq
where $T_f$ and $T_0$ denote the final and initial core 
temperatures. The first term in the bracket on the right hand side 
corresponds  to cooling due to classical gas of the ions and the 
second term corresponds to the contribution coming from the nonrelativistic 
Fermi sea. The latter  is sub-dominant in the range of final 
temperatures we are  interested in (the factor Z in front of this 
term is due to  $Z$ electrons per ion). Since  
$T_ f\ll T_0$, the age of a dwarf star typically  doesn't depend on 
the initial temperature.  Neglecting the fermion contribution, we find time 
that is needed to cool down to critical temperature $T_f=T_c$
\beq
t_{He}=\frac{3}{5}\frac{k_BT_c M}{Am_uL(T_c)}\simeq (0.76 - 7.6)~\text{Gyr}\,.
\label{ages}
\eeq
Where an order of magnitude interval in (\ref {ages}) 
is due to the interval in  the envelope  metallicity 
composition given in (\ref {eq01}).  We also find  
the corresponding luminosities
\beq
L(T_c) \simeq(10^8~erg/s)\frac{M}{M_\odot}
\left ( \frac{T_c}{\text{K}}\right )^{{7/2} }\simeq 1.5\cdot 
(10^{-4}-10^{-5}) L_{\odot}\,,
\eeq
which are in the range of observable luminosities ($L_{\odot}\simeq
3.84\cdot 10^{33} ~erg/s$).

After the condensation, specific heat of the system dramatically drops 
as the collective excitations of the condensed nuclei become 
massive and  ``get extinct''.  A contribution from the Fermi sea, which is 
strongly suppressed by the value of Fermi energy, becomes the dominant one.   
The phase transition itself would take some time to 
complete, and the drop-off in specific heat will not be instantaneous.
During that time a heat-transfer from the bosonic sector 
to the fermionic one will take place, but this will only 
change temperature of the fermions by a factor of $(1+1/Z)$, 
which,  given our approximations,  can be ignored. 
Effects of  finite duration of the phase transition  
on the shape of the luminosity function will be discussed in the next 
section.  Here, for simplicity we approximate the transition to be  
instantaneous, and retain only the fermion 
contribution to specific heat below $T_c$. Then, the expression for the age of 
the star for $T_f<T_c$, reads as follows
\beq
t_{He}'=\frac{k_B}{CAm_u}\left [ \frac{3}{5}(T_c^{-\frac{5}{2}}-
T_0^{-\frac{5}{2}})+Z\frac{\pi^2}{3}\frac{k_B}{E_F}
(T_f^{-\frac{3}{2}}-T_0^{-\frac{3}{2}})\right ].
\label{eq02}
\eeq
Notice the difference of (\ref {eq02}) from  (\ref {eq03}) -- 
in the former $T_f <T_c$ and it is $T_f$ that enters as final temperature 
in the  fermionic part, while  $T_c$ should be taken as the 
final temperature in the bosonic part. 

\begin{figure}[htp]
\centering
\includegraphics{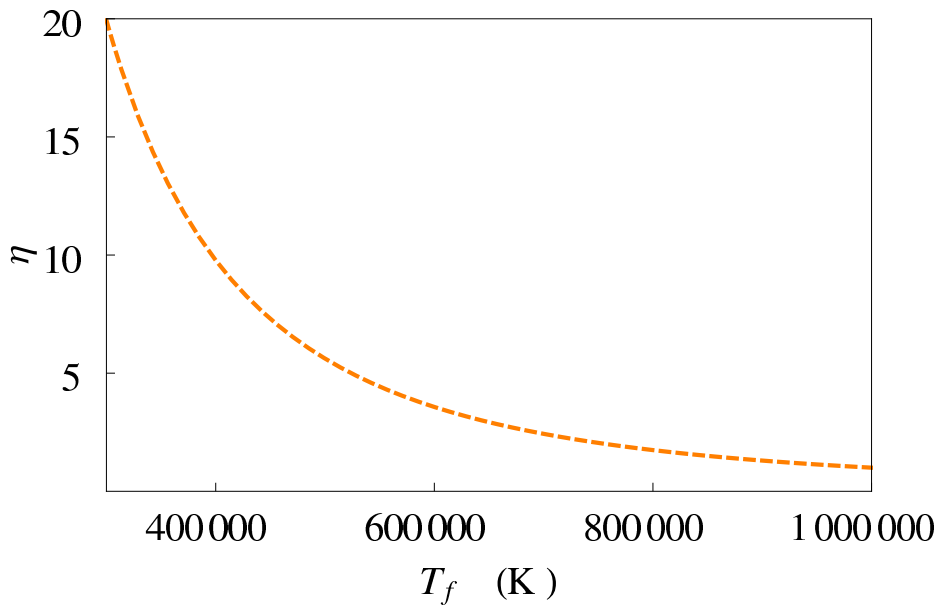}
\caption{The ratio of the ages, as a function of final temperature, for 
two identical helium dwarf stars,  with and without the  
interior condensation.}
\end{figure} 

Figure 1 gives the ratio of ages, $\eta= {t_{He}/ t_{He}'}$, 
for two identical helium dwarf stars, with and without the 
interior condensation. The considered range of final temperature 
is $(3\cdot10^5- 10^6)~K$. At the upper edge  of this interval,
 the luminosity of the star, depending on the metallicity interval given in 
(\ref {eq01}),   is $(10^{-4}- 10^{-5}) L_{\odot}$, while  
for the lower edge of the temperature interval we get
$2 \cdot ( 10^{-6} - 10^{-7}) L_{\odot}$.
As is seen from this figure, the charged condensation substantially 
increases the rate of  cooling. At the lower end of the considered 
temperature  interval  the age is about twenty times less than it 
would have been without the condensation phase.

With the decrease of density, the critical and Debye temperatures 
drop as $T_c\propto \rho^{\frac{2}{3}}$ and 
$\theta_D\propto\rho^{\frac{1}{2}}$ respectively, whereas the 
crystallization temperature drops only as 
$T_{cryst} \propto \rho^{\frac{1}{3}}$. Hence, the 
critical and  crystallization temperatures would become equal 
for  the dwarfs of low density. This would happen for densities 
$\sim 2\cdot  10^{4}~g/cm^3$, which are very low and unlikely to be present 
in white dwarfs cores.  Hence, we'd expect that all He WDs 
should  have a large fractions of their cores in  
the condensed state.

To summarize, the charged condensation of white dwarf interiors 
would have a significant impact on cooling curves of helium 
dwarf stars at the faint end of the luminosity spectrum. 
Starting from a certain temperature scale, $\sim T_c$ , the 
ages of the stars become effectively "frozen", whereas the 
temperature and luminosity decrease rather fast. Such a change 
in the cooling regime would leave its signatures on the luminosity 
function of helium white dwarfs, which we'll  discuss below.

\subsubsection*{Luminosity Function}

The luminosity function (LF) $\phi(\frac{L}{L_\odot})$ 
can be used to  test the white dwarf 
evolution  models against  the observational data; it 
is defined  as follows: 
\beq
\phi(\frac{L}{L_\odot})d(\log(\frac{L}{L_\odot}))\equiv
\text{number-density of WD's per unit interval of}~(\log(\frac{L}{L_\odot})).
\eeq
Under the assumption of the standard star formation rate, uniform 
both in space and time, the luminosity function takes 
a simple power-law form 
\beq
\phi\propto\left 
[ \frac{d\log(\frac{L}{L_\odot})}{dt}\right ]^{-1}
\propto L^{\frac{n(k+1)}{4}-1},
\eeq 
where the constants $k$ and $n$ are the exponents defining 
temperature dependence of specific heat and luminosity:
\beq
c_v\propto T^k, \qquad L\propto T^{\frac{4}{n}}.
\eeq
For instance, $n$ equals to ${8/7}$, once the  
Kramer's opacity is adopted for the description of 
the dwarf atmosphere. The log of the LF corresponding to 
cooling in the Mestel regime 
($c_v=3k_B/2$), is just a line with a slope  equal to ${5/7}$. 

\begin{figure}[htp]
\centering
\includegraphics{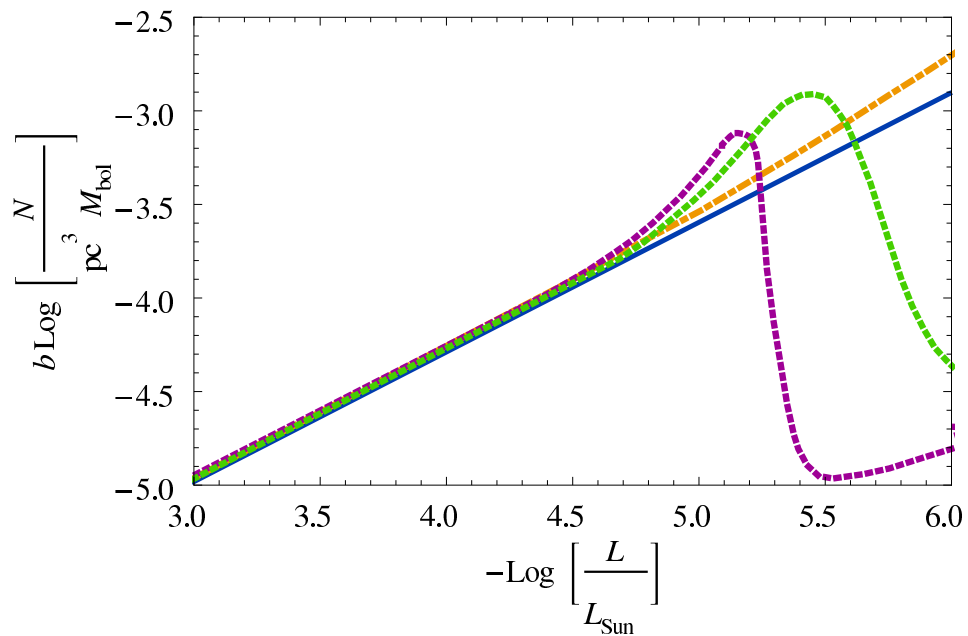}
\caption{A schematic sketch of the luminosity function for 
helium-core  white dwarfs. The absolute normalization of the function is 
set by the constant $b$ which is determined by their formation rate. 
The blue line represents the Mestel regime. 
The shape of the luminosity function near the condensation point depends 
on the details of the corresponding phase transition.}
\end{figure}

The condensation of the core would induce significant deviations from the 
Mestel curve for helium white dwarfs. The exponential suppression 
of specific heat of the ions in the condensate 
doesn't occur instantaneously. Typically, for a quantum liquid, 
there is a transition 
region of width $\sim \Delta T_c$, in which the specific heat changes 
in a certain way.  Therefore, starting from some temperature 
$T_c - \Delta T_c$, at which the specific heat of the 
ions can be neglected and a stage of rapid cooling begins 
(the age of the star ``freezes''),  the luminosity function starts to 
drop dramatically (Fig. 2).  The stars, previously belonging to the 
same logarithmic luminosity bin, will spread among 
a few bins as a result of suddenly accelerated cooling.

Not much can be said with certainty of $\Delta T_c$ and the 
temperature dependence of specific heat near the 
phase transition point. In a well-known case of quantum liquid of 
neutral helium atoms  for instance, the specific heat exhibits the 
so-called ``Lambda''  behavior.  In analogy with the latter, 
one may expect the specific  head to increase near the phase transition 
point before  dropping dramatically below $T_c$. If so, the parameter 
$\Delta T_c$ would  give  the width of the transitional  region.  
Depending on the details of the function $c_v(T)$ near the phase 
transition point, the luminosity function of He WDs could  take  
different shapes near its maximum, as it is schematically illustrated 
in Fig. 2.

What is independent of the above uncertainties, however, is the 
fact that the LF will experience a significant drop-off after the 
charged condensation phase transition is complete. This is due to 
the ``extinction''  of the bosonic quasiparticles below the phase 
transition point. In fact, the LF function will drop by a factor of 
$\sim 200$,  which would be reflected as a drop-off  by about two units of the 
vertical axis from its maximum shown on Fig 2. After the drop-off the 
log of the LF function starts growing again but now 
with the slope equal to $3/7$ --  due to the specific heat of the electrons. 
The latter part of the LF  is shown only for one of the curves on Fig 2.
Whether this drop-off can explain the termination of the 
24 He WD sequence found in \cite {24}, 
remains to be seen in more detailed studies.

\subsubsection*{Superdense Carbon White Dwarfs}

Finally, we briefly mention another possible subclass of white dwarf stars, 
that could undergo the core condensation. These are superdense 
dwarfs composed mainly of carbon, with 
masses  close to the Chandrasekhar limit $\sim 1.4 ~M_\odot$, 
and central densities $\sim 10^{10} g/cm^3$. 
Such a star would be very close to the neutronization 
threshold ($5\cdot  10^{10} g/cm^3$ for carbon nuclear matter), 
and it's   
critical temperature $T_c\sim 8\cdot10^7~K$ would be 
greater than the crystallization temperature 
$T_{\text{cryst}}\sim 4 \cdot10^7~K$,  making condensation 
a possibility\footnote{At these high temperatures the WD 
cooling rate is significantly affected by the neutrino emission. 
Our goal,  however, is to identify qualitative differences 
of  cooling in the condensate and crystalline phases. 
For this purpose, and for simplicity, we ignore the neutrino 
contributions, which should certainly be 
taken into account for these WDs in more precise studies.}.

Assuming the helium dominated envelope 
($Y$ denotes the helium fraction) with a 
$10\%$ metallicity
\beq
X\simeq 0,\qquad Y\simeq 0.9, \qquad Z_m\simeq 0.1,
\eeq   
we find the cooling time needed to reach down to 
temperature $T_c \sim 8\cdot10^7~K$ 
\beq
t_{C}=\frac{3}{5}\frac{k_BT_c M}{Am_uL}\simeq 3\cdot 10^{-3}~\text{Gyr}.
\eeq
This temperature corresponds to luminosity $L\simeq 3.4 L_\odot$.
Once the star enters the condensation regime, it cools faster. 
For instance, it would reach the luminosity $L\simeq 10^{-6} L_\odot$ in 
$t=0.7~{\rm Gyr}$ 
after the condensation. On the other hand, if one assumes that 
this star crystallizes instead of condensing its core, then it would reach the 
crystallization point in $10^{-2}~\text{Gyr}$, while it would 
take only  $\sim 0.1~{\rm Gyr}$ longer for this star to  
cool down  to  $L\simeq 10^{-6} L_\odot$, as compared with the 
condensation case.  In the crystallized phase for 
this star the cooling due to the fermion contributions is significant.  
Therefore, the log of the LF for such a star
would not have a sharp drop-off; instead, after the phase transition 
the LF would just change its slope to $3/7$ in the low luminosity region.

We also note that for dwarf stars with relativistically degenerate cores, 
the polytrope models  yield more dramatic density profiles  -- density in 
the center exceeds  the average density by a factor of $\sim  50$.   Since 
the crystallization/critical  temperature equality is achieved at 
densities $\sim  10^9~g/cm^3$ for carbon interiors, we should not expect 
the entire core to condense.  In realistic calculations of cooling rates 
of  the superdense carbon WDs this should also be taken into account.

Magnetic properties  of condensed WDs, and their similarity to 
type II supeconductors, will be discussed in \cite {GGRRmag}.

Finally, we mention that charged condensation may also take place 
in some other astrophysical  objects where the densities and 
temperatures are appropriate. Crusts of neutron stars may  
be a place to look at.  

\vspace{0.2in}

\begin{center}
{\bf   Acknowledgments}
\end{center}

We'd like to thank Paul Chaikin, Daniel Eisenstein, 
Leonid Glazman, Andrei Gruzinov, Andrew MacFadyen, Aditi Mitra, 
Slava Mukhanov, Rachel Rosen and Malvin Ruderman for useful
discussions and correspondence. GG  was supported by  the NSF 
and NASA grants PHY-0758032, NNGG05GH34G. DP acknowledges the NYU 
James Arthur graduate fellowship support. 

\vspace{0.2in}

\section*{Appendix}

For purposes of studying their cooling, white dwarfs are 
well described by a simple 2-component model. The core consists 
of the ion gas and degenerate electrons with large mean free path 
and heat conductivity, making it almost isothermal. The cooling 
takes place through the non-degenerate surface layer,  the envelope, 
that 
surrounds the core. The photon diffusion equation, describing 
the energy flow from the core to the outer layers, has the following form
\begin{equation}
L=-4\pi r^2 \frac{c}{3\kappa\rho}\frac{d}{dr}(aT^4),
\end{equation}
where $L$ is luminosity, $aT^4$- energy density of a blackbody, 
$\kappa$ is the opacity of the stellar matter and 
$\rho$-mass density. We can rewrite the last equation in the following form
\begin{equation}
\frac{dT}{dr}=-\frac{3}{4ac}\frac{\kappa\rho}{T^3}\frac{L}{4\pi r^2}.
\label{eq2}
\end{equation}
For the opacity, we use Kramer's approximation 
$\kappa=\kappa_0\rho T^{-3.5}$. 

The equation of the hydrostatic equilibrium (where $P$ denotes the pressure 
and $m(r)$ - the mass inside a sphere of radius $r$)
\begin{equation}
\frac{dP}{dr}=-\frac{Gm(r)\rho}{r^2},
\label{eq1}
\end{equation}
combined with  the equation of state for a classical ideal gas
\begin{equation}
P=\frac{\rho}{\mu m_u}k_BT\,,
\label{eq4}
\end{equation}
(where $\mu$ and $m_u$ are the mean molecular weight and 
atomic mass unit respectively), can be  integrated 
with the boundary condition $P=0$ at $T=0$ to obtain
\begin{equation}
\rho=\left ( \frac{2}{8.5}\frac{4ac}{3}\frac{4\pi GM}
{\kappa_0 L}\frac{\mu m_u}{k_B}\right )^{\frac{1}{2}}T^{3.25},
\label{eq3}
\end{equation}
where $\kappa_0 = 4.34 \cdot 10^{24}Z_m(1+X)$ $cm^2/g$ is the 
commonly used value for the opacity constant, $X$ is the mass 
fraction of hydrogen,   and $Z_m$  - that of heavy elements 
(all elements except hydrogen and helium), while $M$ is the mass of 
the star.  Equation (\ref{eq3}) works as long  as we deal with 
nondegenerate matter. To estimate the limits of its applicability, 
we should equate the pressure of nondegenerate \emph{electrons} 
(obtained from~(\ref{eq4}) by replacing $\mu$ by mean molecular 
weight per electron $\mu_e$) to that of the nonrelativistic 
electron gas in the outer layers of the core
\begin{equation}
\frac{\rho_{\ast} k_BT_{\ast}}{\mu_e m_u}=
{3^{\frac{2}{3}} \pi^{\frac{4}{3}} \over 5}
{\hbar ^2 \rho_{\ast}^{{5/3}} \over m_e (m_u \mu_e)^{5/3}}\,,
\label{eq5}
\end{equation}
where $\rho_{\ast}$ and $T_{\ast}$ are the density and 
temperature of the star at the core-surface boundary. 
Combining equations (\ref{eq5}) and (\ref{eq3}), we obtain a  
useful relation for the dependence of the luminosity on the temperature 
at the core-surface boundary (which, due to the isothermality of 
the core, can be used as a good characteristic of the core temperature) 
\begin{equation}
L=\gamma\,\frac{\mu}{\mu_e^2}\frac{1}{Z_m(1+X)}\frac{M}
{M_{\odot}}(T_{\ast})^{\frac{7}{2}}\equiv CM(T_{\ast})^{\frac{7}{2}}\,,
\label{eq6}
\end{equation}
where $\gamma\simeq 5.7 \cdot 10^5 ~erg/s$ is a universal constant, 
and $C$ is a constant that depends on the chemical composition of 
the envelope and varies from a star to star. A more general form of 
the  last equation may be written as follows:
\beq
L\propto T^{\frac{4}{n}},
\eeq
where $n$ is a constant, equal to $8/7$ in case of the envelope  
with Kramer's opacity.

\end {document}